\documentclass[11pt, draftclsnofoot, onecolumn]{IEEEtran}

\usepackage{cite,graphicx,amsmath,amssymb}
\usepackage{subfigure}
\usepackage{fancyhdr}
\usepackage{mdwmath}
\usepackage{mdwtab}
\usepackage{balance}
\usepackage{xcolor}
\usepackage{bm}
\usepackage{amsthm}
\usepackage{algorithm}
\usepackage{algorithmic}
\usepackage{multirow}
\usepackage{flafter}
\usepackage{enumerate}

\begin{document}
\title{Multiple Transmit Power Levels based NOMA for Massive Machine-type Communications}

\author{
\IEEEauthorblockN{Wenqiang Yi, Wenjuan Yu, Yuanwei Liu, Chuan Heng Foh, Zhiguo Ding and Arumugam Nallanathan}

\thanks{W. Yi, Y. Liu, and A. Nallanathan are with the School of Electronic Engineering and Computer Science, Queen Mary University of London, London, E1 4NS, U.K. (Email:\{w.yi, yuanwei.liu, a.nallanathan\}@qmul.ac.uk)}
\thanks{W. Yu is with the School of Computing and Communications, InfoLab21, Lancaster University, Lancaster, LA1 4WA, U.K. (Email: w.yu8@lancaster.ac.uk) }
\thanks{C. Foh is with 5G Innovation Centre, Institute for Communication Systems, University of Surrey, Guildford, GU2 7XH, U.K. (Email: c.foh@surrey.ac.uk)}
\thanks{Z. Ding is with School of Electrical and Electronic Engineering, University of Manchester, Manchester M13 9PL, U.K. (Email: zhiguo.ding@manchester.ac.uk)}
}
\maketitle

\begin{abstract}
  This paper proposes a tractable solution for integrating non-orthogonal multiple access (NOMA) into massive machine-type communications (mMTC) to increase the uplink connectivity. Multiple transmit power levels are provided at the user end to enable open-loop power control, which is absent from the traditional uplink NOMA with the fixed transmit power. The basics of this solution are firstly presented to analytically show the inherent performance gain in terms of the average arrival rate (AAR). Then, a practical framework based on a novel power map is proposed to associate a set of well-designed transmit power levels with each geographical region for handling the no instantaneous channel state information problem. Based on this framework, the semi-grant-free (semi-GF) transmission with two practical protocols is introduced to enhance the connectivity, which has higher AAR than both the conventional grand-based and GF transmissions. When the number of active GF devices in mMTC far exceeds the available resource blocks, the corresponding AAR tends to zero. To solve this problem, user barring techniques are employed into the semi-GF transmission to stable the traffic flow and thus increase the AAR. Lastly, promising research directions are discussed for improving the proposed networks.
\end{abstract}
\begin{IEEEkeywords}
Average arrival rate, massive machine-type communications, non-orthogonal multiple access, power map, semi-grant-free transmission, user barring
\end{IEEEkeywords}

\section{Introduction}
With the explosive of Internet-enabled devices in various new fields, e.g., medicine, agriculture, industry, etc., massive machine-type communications (mMTC) become one of the main use cases for the next generation of cellular networks with Internet-of-things (IoT)~\cite{Intro1}. In contrast to human-type communications, mMTC commonly has a fixed and low data rate in the uplink transmission. Although the number of devices is huge, only a part of these devices will be activated to transmit their data at each time slot. Therefore, it is intractable to allocate resources to all devices in advance~\cite{Intro2}. To solve this problem, the limited spectral resources in mMTC networks need at least two new abilities, i.e., random access and multiplexing. Designing advanced techniques for providing massive connectivity becomes a burning problem for upcoming mMTC era~\cite{Intro3}.

By introducing a new freedom degree, namely the power domain, non-orthogonal multiple access (NOMA) is able to serve more than one user in the traditional time/frequency resource block (RB)~\cite{Intro4}. For uplink-NOMA, users select one RB at random or based on a designed policy to upload their data. At the base station (BS) end, the successive interference cancellation (SIC) technique is employed to decode signals by mitigating intra-RB interference. Since the traditional uplink-NOMA with the fixed transmit power need closed-loop power control to find the optimal power for each user, it is not suitable for mMTC with massive potential transmitters due to introducing additional overhead. To enable open-loop power control, the BS broadcasts a set of transmit power levels (TPLs) for all IoT devices to choose from, which is capable of increasing both the connectivity and spectral efficiency~\cite{Intro6}.

Motivated by the aforementioned advantages of multiple transmit power levels based NOMA (MTNOMA), this work aims to offer a tractable method to integrate MTNOMA into mMTC for enhancing the connectivity. From the perspective of the received power levels (RPLs), we first introduce the basics of MTNOMA-mMTC networks and evaluate the connectivity gain. To explore the mapping between the RPLs and TPLs, we introduce a mathematical model with the aid of stochastic processes and statistical channel models. Under this mathematical model, we create a power map to allocate a set of TPLs to each geographically divided region to reduce traffic collisions and energy consumption. No instantaneous channel state information (ICSI) but some relatively constant geographical information is needed for this practical framework. After that, we select semi-grant-free (semi-GF) transmissions for the considered MTNOMA-mMTC networks since it combines the safety of grant-based (GB) transmissions and the flexibility of GF transmissions. When the number of active IoT devices further increases, the connectivity of semi-GF transmission gradually decreases. To release the potential of the semi-GF transmission under this case, we employ user barring techniques to limit the number of simultaneously transmitting GF devices. Finally, the proposed insights are summarized in the conclusion and several promising research directions are provided as well.

\section{MTNOMA-mMTC Basis}
\begin{figure}[t!]
  \centering
  \includegraphics[width=1 \linewidth]{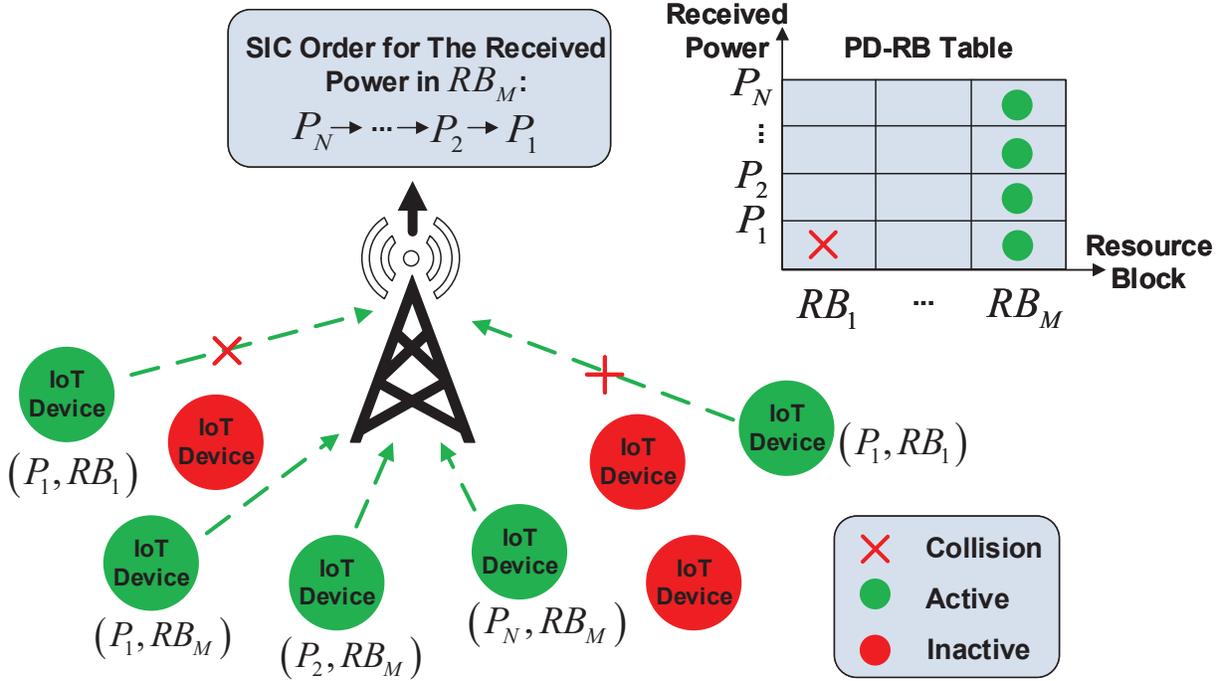}
  \caption{Illustration of MTNOMA-mMTC networks, where the BS broadcasts the PD-RB table to all IoT devices for randomly choosing. In each orthogonal RB, the SIC is employed to decode the signals of active IoT devices. When more than one IoT devices select the same PD-RB, there is a collision.}\label{fig1}
\end{figure}

As illustrated in Fig.~\ref{fig1}, we discuss a single-cell MTNOMA-mMTC network from the perspective of RPLs, where massive intermittently active IoT devices upload information to the serving BS via $M$ orthogonal (time/frequency) RBs. To enable NOMA, each RB is divided into $N$ RPLs. The combination of RPLs and RBs $(P_i,RB_j)$ is defined as power-domain resource blocks (PD-RBs), where $i = \{1,..,N\}$ and $j = \{1,...,M\}$. In this uplink transmission, the BS first broadcasts the PD-RB table to all IoT devices. Then, these IoT devices independently select one PD-RB to transmit their data. After that, the BS attempts to decode the received signal in each RB according to SIC techniques, which has the following process: 1) The BS decodes the signal with the strongest RPL; 2) The decoded signal is cancelled from the original signal; and 3) The BS moves on the next signal with the second strongest RPL. Based on this principle, the decoding order is the same with the strength order of RPLs. To ensure the success of SIC processes with a target signal-to-interference-plus-noise ratio (SINR), the ratio of an RPL to the summation of its lower RPLs plus noise should be larger than this target SINR. It is worth noting that when multiple devices choose the same PD-RB, a collision occurs. This collision results in failed decoding for the devices in the same and lower PD-RBs\footnote{Comparing with a certain PD-RB with an RPL $P_i$, the lower PD-RBs represents the PD-RBs in the same RB whose RPLs are smaller than $P_i$, and vice verse.}, while for the devices in the higher PD-RBs, there is still a success probability. For example, supposing that one device at $(P_N,RB_1)$, two devices at $(P_{N-2},RB_1)$, and no device at $(P_{N-1},RB_1)$, the devices from $(P_{N-2},RB_1)$ to $(P_1,RB_1)$ cannot be decoded, but the device at $(P_N,RB_1)$ still has a successful uplink transmission. As a conclusion, when a collision happens in one RB, MTNOMA-mMTC systems still have a probability to partially decode the received signal in this RB.

One major challenge for mMTC networks is to provide massive access within limited spectrum resources. To this end, the average number of successful access in each RB, namely the average arrival rate (AAR), should be increased. Comparing with the conventional orthogonal multiple access (OMA) scenarios, the considered MTNOMA-mMTC has at least two advantages: 1) For MTNOMA-mMTC, the available access resources, i.e., PD-RBs, in each RB are $N$ times larger than those for OMA-mMTC; and 2) For collision-existed RBs, MTNOMA-mMTC has a probability to decode partial information, while OMA-mMTC has no uplink throughput. Therefore, we are able to conclude that MTNOMA-mMTC outperforms OMA-mMTC in terms of the connectivity. It is worth pointing out that the AAR gain does not increase linearly with $N$ because of the nonorthogonality between PD-RBs. One drawback for applying MTNOMA-mMTC is the transmit power of devices is enlarged to combat extra NOMA-introduced interference. Fortunately, the energy consumption may be still less than OMA-mMTC as higher connectivity contributes to fewer attempts for retransmissions. As a result, there is a tradeoff between low energy consumption and high connectivity for MTNOMA-mMTC networks.

This basis analysis is based on RPLs for simplicity. However, comparing with RPLs, it is more practical for IoT devices to know their TPLs, which helps to avoid additional computation cost. Due to the huge amount of IoT devices and uncertain channel state, it is intractable to acquire ICSI of all devices at the BS end. The mapping between RPLs and TPLs is one challenging problem. Moreover, letting the IoT devices with weak channel gain to choose the high PD-RBs reduces energy efficiency. How to allocate PD-RBs to IoT devices becomes another challenging problem. We propose a practical framework in the next section to address these two problems.

\section{Practical Framework with Power Map}
\begin{figure}[t!]
  \centering
  \includegraphics[width=1 \linewidth]{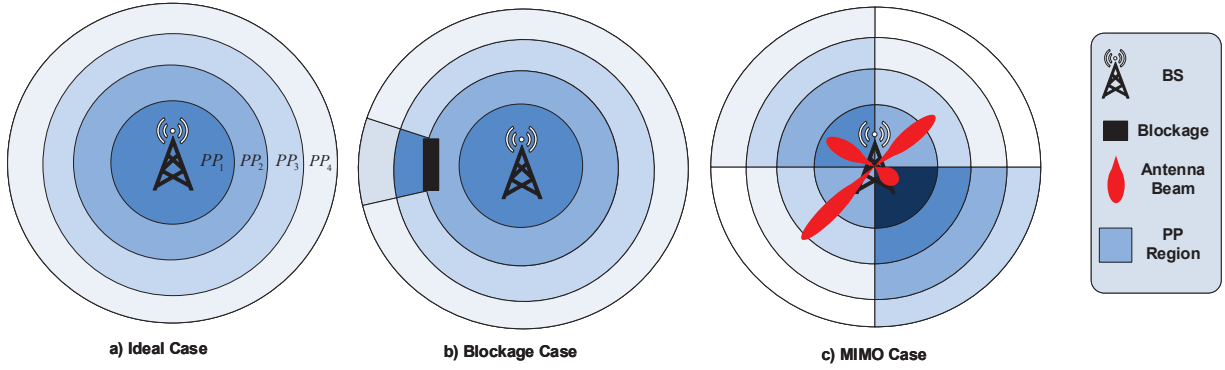}
  \caption{Illustration of the considered power map: a) an ideal case with the omnidirectional antenna pattern and no blockage; b) a blockage case with the omnidirectional antenna pattern and an obstacle; and c) a multi-input and multi-output (MIMO) case with no blockage but different antenna gain for each direction. The average TPL in PPs obeys PP$_1 > $PP$_2>$PP$_3>$PP$_4$.}\label{fig5}
\end{figure}

The power allocation in terms of TPLs for MTNOMA-mMTC is difficult since BSs commonly have only partial or even no ICSI of active IoT devices, especially at the beginning of the transmission. Without a closed-loop power control, a global resource management is intractable under this case. However, the closed-loop power control introduces additional traffic loads, which deteriorates the connectivity. Note that RPLs for MTNOMA-mMTC networks are mainly decided by the transmit power and the corresponding channel gain. For a certain environment, only the transmit power is controllable. Fortunately, the channel gain can be estimated via the geographical locations of IoT devices and practical statistical channel models~\cite{path_loss}. Based on this idea, we propose a mathematical model to connect RPLs with TPLs. With the aid of this model, a part of all available TPLs is selected for each geographically divided region to limit power consumption~\cite{JChoi20171}, which form a power map. The set of the selected TPLs is called a power pool (PP) and each PP corresponds to a part of PD-RBs in an RB. Since this power map is generated via stochastic processes and statistical models, it can be designed before the actual runtime, which enables an open-loop power control. The ICSI is not necessary for the power map design but it is important for performance improvement. In other words, the mathematical framework based power map has limited accuracy, but if the ICSI of IoT devices can be acquired during the later transmission, we are able to improve the accuracy by updating the original power map.

Now, the main question is how to design this power map? The first step is to divide the considered area into several regions based on pre-determined RPLs. This division needs to consider at least two factors: blockage and antenna gain. As shown in Fig.~\ref{fig5}, comparing with the ideal case with omnidirectional antenna pattern and no blockage, the PPs for the regions behind the blockage and those experiencing small antenna gain need to be high\footnote{If the average TPL for PP$_1$ is higher than PP$_2$, we define that PP$_1$ is higher than PP$_2$ and vice verse.}. For a certain application, e.g., factories, the first step can be achieved by field measurements to enhance accuracy. The second step is to design the PP for each region. Since this design needs the learning ability to update the original power map based on the ICSI of IoT devices, machine learning (ML) becomes a promising method for solving it. On the one hand, ML is able to solve NP-hard optimization problems, most of which cannot be handled via the traditional optimization methods. On the other hand, ML updates its parameters based on old experience and new environments, which matches the considered PP design. For long-term communications, the correlation between different decisions cannot be ignored, so reinforcement learning (RL) that learns the best policy (a sequence of actions) becomes a reliable choice.

Note that deep reinforcement learning (DRL) is one of the RL algorithms. By applying deep neural network (DNN), the Q-function of DRL $Q(s,a,\theta)$ has one more parameter, i.e., $\theta$, than that of the traditional Q learning $Q(s,a)$, where $s$ and $a$ represent states and actions, respectively. This new parameter introduces the prediction ability to estimate the Q-value for the states that do not appear before. Instead of memorizing all combinations of states and actions, DRL only needs to store the set of $\theta$ which has a smaller size. Therefore, DRL requires a smaller memory and converges faster than the Q learning~\cite{DRL}. We use a simple multi-agent DRL to design the PP for long-term communications. The designs of states, actions, and rewards are listed as follows:
\begin{figure}[t!]
  \centering
  \includegraphics[width=1 \linewidth]{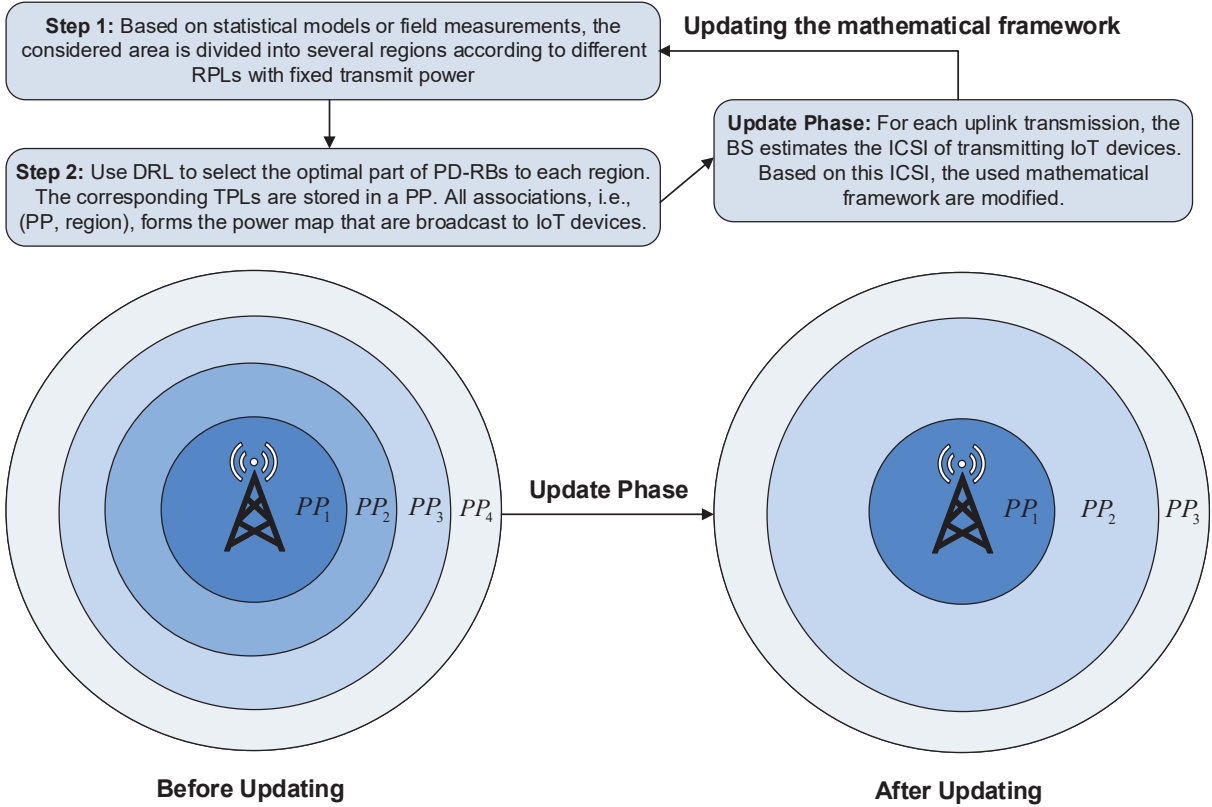}
  \caption{The design process of the power map. After updating the applied mathematical framework based on the ICSI, both the region division and the PP can be modified to increase the performance gain.}\label{fig6o}
\end{figure}
\begin{itemize}
  \item \textbf{State Space:} Each IoT device represents one agent and it interacts with the wireless environment. Every SINR for decoding is a state, so the size of state space is equal to the number of active IoT devices.
  \item \textbf{Action Space:} Each agent randomly chooses one RP-RB to transmit. Therefore, every action corresponds to a certain combination of RPLs and RBs. The size of the action space for each agent is $M\times N$.
  \item \textbf{Reward:} For different optimization problems, the reward can be the corresponding objective function. For example, if we aim to minimize power consumption, the states are first converted to power consumption. When the current state is smaller than the previous state, the agent is rewarded with the reciprocal of this power consumption. Otherwise, the reward is zero. A centralized reward method can be applied to combat selfish manners, so all agents receive the same reward.
\end{itemize}
The training details of DRL is provided in~\cite{DRL2} and hence we omit it here. After convergency, the optimal TPLs located in each region are stored in a PP. The power map design can be summarized in Fig~\ref{fig6o}. After receiving the broadcast information from the BS, IoT devices randomly choose one TPL from the location corresponded PP to send its data. Based on this power map aided framework, the next job is to design the transmission scheme for MTNOMA-mMTC networks.

\section{Semi-Grant-free Transmission}
Although most IoT devices can tolerate retransmissions, there are still some primary IoT devices requiring undistracted transmission. Therefore, the GB transmission cannot be ignored in MTNOMA-mMTC networks. Unlike human-type communications that focusing on high-speed transmission, the majority applications of mMTC need a low data rate but high connectivity. Therefore, the use of the traditional GB transmission offers more capacity than need for mMTC scenarios, which results in resource waste, especially in the power domain. Note that by removing uplink scheduling requests and dynamic scheduling grants, GF transmission is able to provide massive connectivity for short-packet communications in an arrive-and-go manner~\cite{Liang2}. To enhance the connectivity of MTNOMA-mMTC, the extra capacity under the GB scheme can be used to provide additional access via GF schemes, which forms semi-GF transmissions~\cite{SGF}. It is noteworthy that comparing with GF transmissions without any reserved RBs, some RBs need to be reserved for GB devices in semi-GF transmissions. In this part, we consider a two-user NOMA as a case study.

\begin{figure}[t!]
  \centering
  \includegraphics[width=1 \linewidth ]{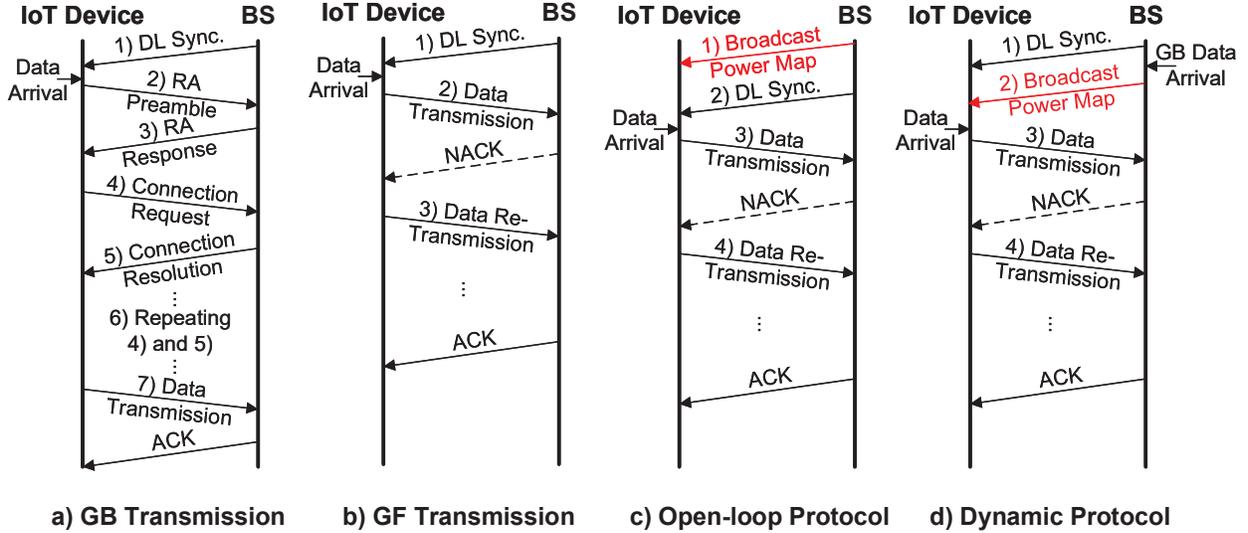}
  \caption{The process comparison between GB, GF, and semi-GF transmission, where DL, Sync., RA, ACK, NACK represent downlink, synchronization, random access, acknowledgement, no acknowledgement, respectively. The red process in semi-GF transmission, namely in sub-figure c) Open-loop Protocol and d) Dynamic protocol, is the new process due to introducing NOMA techniques.}\label{fig2}
\end{figure}

In semi-GF transmission, one GB device connects to the BS in the previous user association process, and then one GF device joins in the same RB to form a NOMA cluster. Comparing with GF schemes, semi-GF needs one more process to share the information of the connected GB device. This process is to ensure the quality of service (QoS) of the GB device the same as that under the conventional OMA case. The detailed steps of the new process is listed as follows:
\begin{itemize}
    \item \textbf{Step 1:} The BS estimates the received signal power of the GB device and calculate an intra-RB \emph{threshold} according to the QoS of the GB device.
    \item \textbf{Step 2:} The BS first deletes the TPLs that break the \emph{threshold requirement} in all PPs and then broadcasts the modified power map to all possible intra-RB GF devices.
    \item \textbf{Step 3:} After receiving the broadcast message, all possible GF devices randomly select one TPL in its own PP for uploading. If GF devices receive a void PP, they keep silence.
\end{itemize}
Based on the SIC order in NOMA clusters, there are two types of thresholds in \textbf{\textbf{Step 1}}. The first type is the \emph{lower-limit threshold}, which is suitable for delay-tolerant GB devices. Under this case, the threshold is the closest PD-RB to the received signal power. The \emph{corresponding threshold requirement} is that only the higher PD-RBs can be used to generate the power map. The advantage of this type is that the GB device has interference-free decoding and no error floor exists in high signal-to-noise (SNR) regions. The disadvantage is that one SIC process is needed before decoding, which results in additional delay. The second type is the \emph{upper-limit threshold}, which is preferable for delay-sensitive GB devices. Under the premise of guaranteeing the QoS, the threshold is the closest PD-RB to the max affordable interference. The \emph{corresponding threshold requirement} is that only the lower PD-RBs can be used to generate the power map. No delay occurs for the GB device under the second type, but the performance of the GB device is limited by the intra-NOMA interference. It is worth noting if no GB device exists, semi-GF transmissions are able to change the lower-limit threshold to zero or the upper-limit threshold to infinity to proceed GF transmissions.

\begin{figure}[t!]
\centering{\includegraphics[width =0.7\linewidth]
{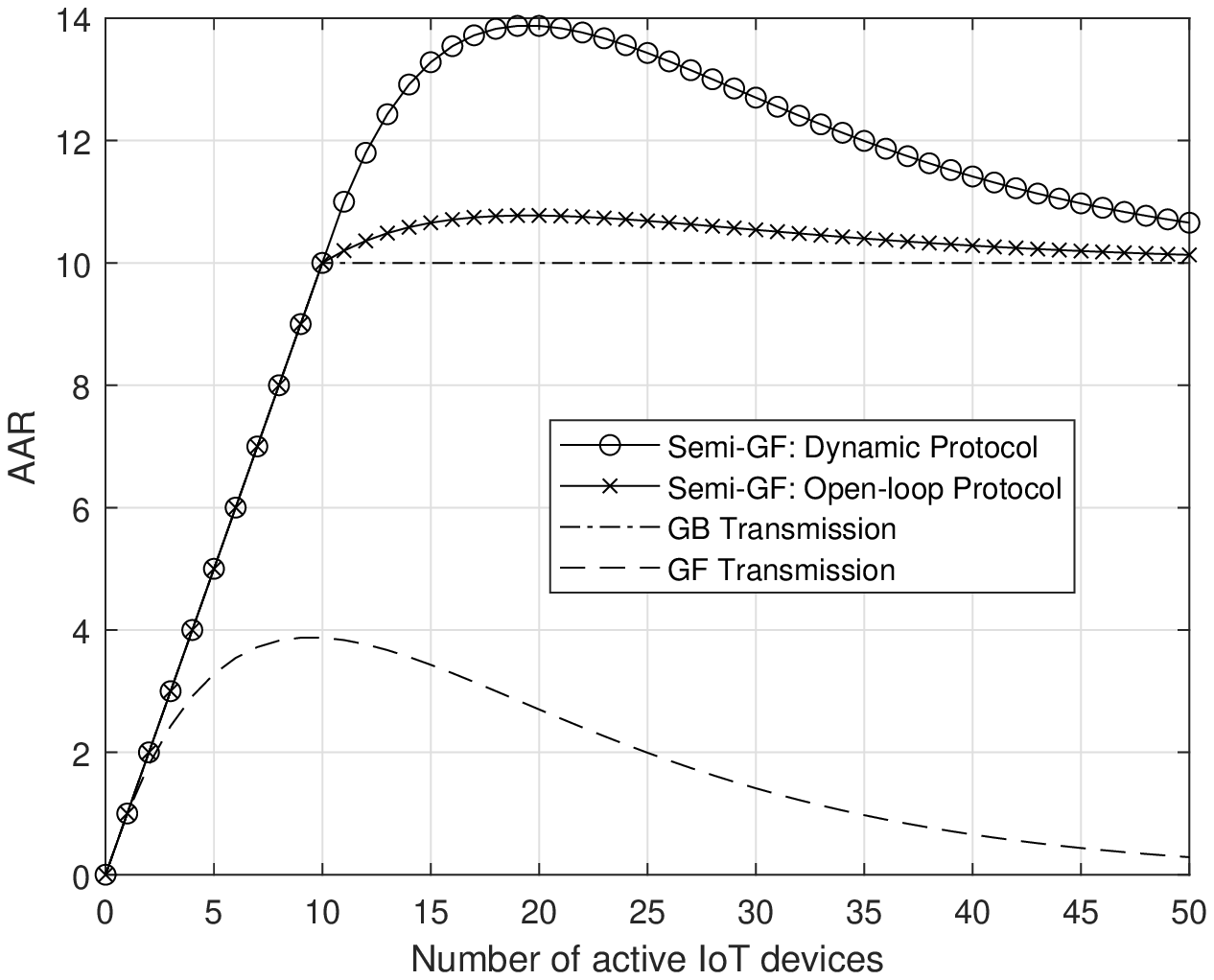}}
\caption{Average arrival rate versus the number of active IoT devices, where $M=10$ and $N=2$. For the open-loop protocol, the probability that the instantaneous signal power of the GB device is lower than the lower-limit threshold (or higher than the upper-limit threshold) is $60\%$. To provide a fair comparison, the semi-GF transmission provides GB channels to all devices when their quantity is less than $10$.}\label{fig3}
\end{figure}

In addition to the threshold, the estimated signal power of the GB device is also controllable. This power can be either the average value or the instantaneous value. Therefore, two semi-GF protocols can be provided~\cite{Semi-GF}: 1) Open-loop protocol; and 2) Dynamic protocol. For the open-loop protocol, the average received signal power of the GB device after a certain estimation period is applied to calculate the threshold. Therefore, outage may happen sporadically. This protocol suits for long-term GB transmissions, e.g., large video uploading. For the dynamic protocol, the instantaneous signal power of the GB device is utilized to figure out the threshold and hence no outage exists. This protocol suits for short-term GB transmission with the strict QoS requirement, e.g., medical sensor communications. The dynamic protocol has lower outage probabilities than the open-loop protocol~\cite{Semi-GF}. Due to using different received power, two protocols have their unique handshakes, which are provided in Fig.~\ref{fig2}, as well as comparing with the traditional GF and GB protocols. Although semi-GF transmission introduces one more process than the traditional GF transmission, it creates a new ability to multiplex the GB spectrum resources, which is a qualitative change from nothing and significantly enhances the spectral efficiency. The AAR comparison is provided in Fig.~\ref{fig3}, when ignoring the overhead, the GB transmission has higher AAR than the GF transmission and the semi-GF transmission outperforms the GB transmission. Due to ensuring the uplink traffic of the GB device all the time, the dynamic protocol has higher AAR than the open-loop protocol.

For serving multiple GF devices in one NOMA cluster, the first type is not a good choice since if a collision happens, the traffic for the GB device cannot be guaranteed. Therefore, we focus on the second type. From Fig.~\ref{fig3}, when the number of active devices increases, the connectivity gain for the semi-GF transmission decreases. To avoid this decline, a traffic stabilization technique is introduced in the next section.

\section{Traffic Stabilization with User Barring Techniques}
Traffic burstiness is a critical issue in semi-GF transmissions, especially when a large number of GF IoT devices are simultaneously activated, e.g., sensors reconnecting after a power outage\cite{MVilgelm2018}. This simultaneous triggering brings in bursty arrivals, which causes lots of collisions and large delays for the collided devices due to subsequent backoff periods. As a result, the connectivity will be significantly degraded. More importantly, the QoS of the GB device cannot be guaranteed under this scenario, which breaks the original intention of the semi-GF design. Therefore, we need to limit the number of active GF devices in each RB. Conventionally, the traffic burstiness issue is alleviated with backoff-based mechanisms or access class barring (ACB) proposed by 3GPP\cite{3GPP22011}. The basic idea of backoff mechanisms is to defer the re-transmissions of collided packets at random. However, the backoff latency may increase exponentially when the number of collided packets increases. The 3GPP ACB technique utilized in LTE-A is proposed for orthogonal radio resources and is used to bar preamble transmissions. Since we aim to address the bursty traffic issue in MTNOMA-mMTC networks, we propose a user barring algorithm for semi-GF transmissions to avoid massive synchronized access demands by redistributing access requests of devices through time~\cite{WYu2020}. It can be applied at the BS to dynamically adjust the barring rate and enable adaptive congestion control based on the real-time traffic load observations.

\begin{figure}[t!]
\centering
\subfigure[Average arrival rate]{\label{Fig6a}\includegraphics[width=0.49\linewidth]{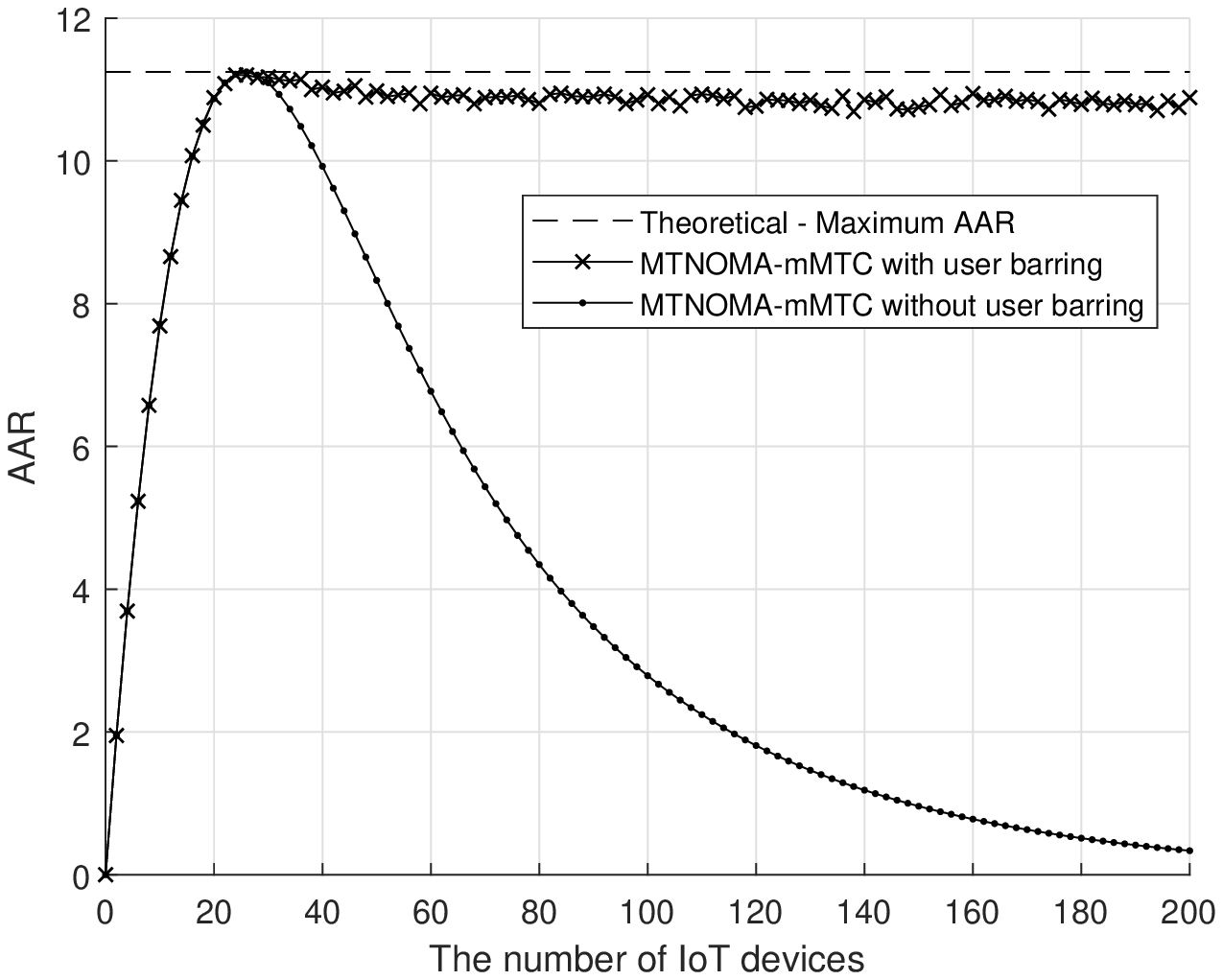}}
\subfigure[Average system Load]{\label{Fig6b}\includegraphics[width=0.49\linewidth]{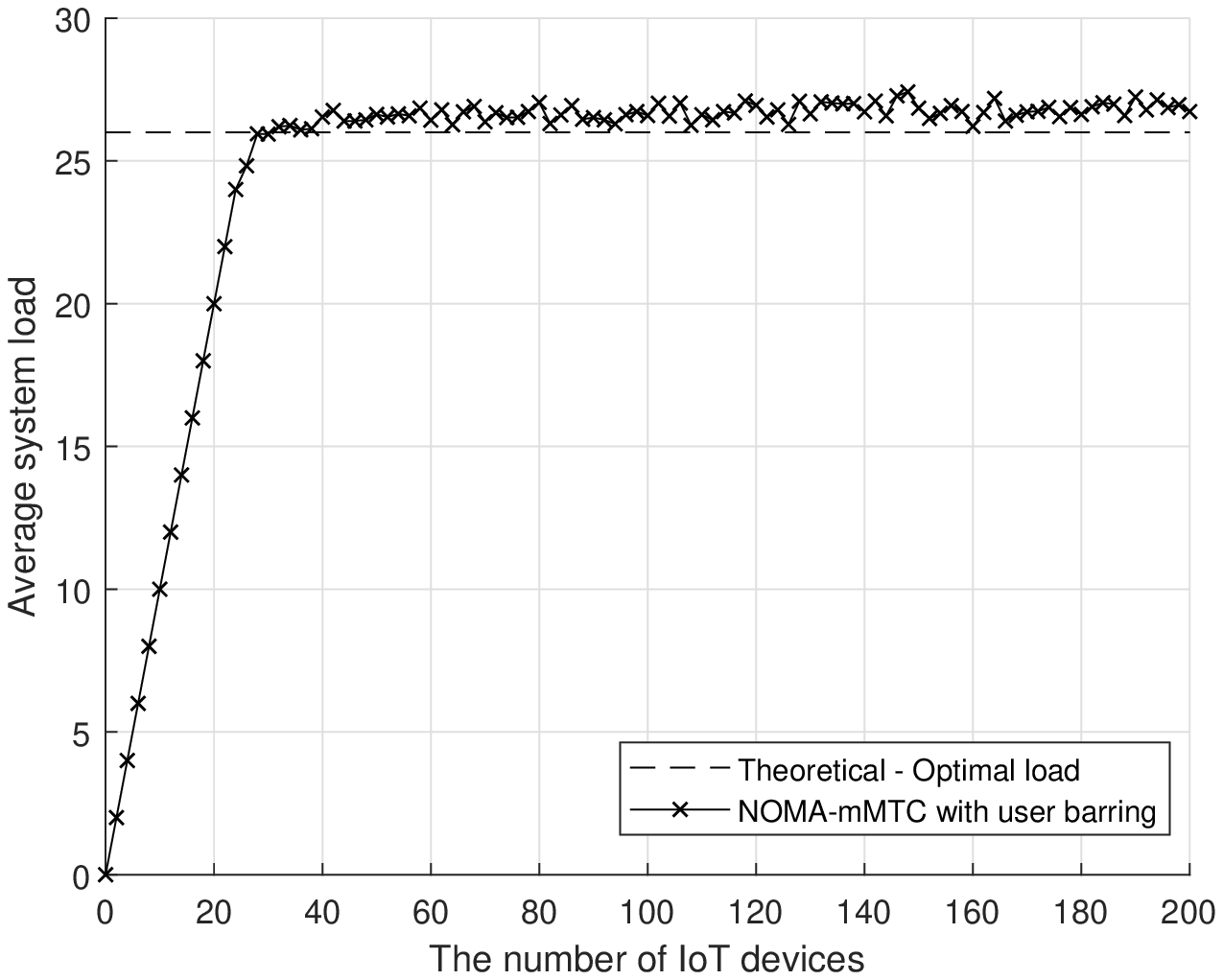}}
\caption{Average arrival rate and average system load versus the number of IoT devices, where M=10 and N=4. All the 4 available RPLs satisfy the constraint that the access of any active devices to each RB will not impact the QoS of GB devices.}
\end{figure}

To better understand the design of user barring in semi-GF transmissions, the detailed steps are given below. Firstly, as noted in Section IV, in order to guarantee the QoS of GB devices, the BS calculates the intra-RB \textit{upper-limit threshold} and generates the power map. Then, the user barring process starts.
\begin{itemize}
    \item \textbf{Initialization:} Given the available TPLs and RBs in the generated power map, the BS uses the analytical expressions in~\cite{WYu2020} to obtain the maximum AAR and optimal load, i.e., the maximum average number of successful access and the optimal number of GF devices in the network.
    \item \textbf{Step 1:} The BS observes the real-time channel outcomes including the instantaneous arrival rate and the number of idle RBs for a fixed barring period. Based on the collected information and the pre-obtained theoretical AAR, the BS estimates the current load in this MTNOMA-mMTC network.
    \item \textbf{Step 2:} Based on the estimated instantaneous load, the BS updates the barring rate for the next barring period, with the aim of maintaining optimal load.
    \item \textbf{Step 3:} The updated barring rate is broadcast to all the associated GF devices, along with the power map. Upon receiving the broadcast information, the GF devices that have packets to send determine their active state by drawing a uniform random value between $0$ and $1$ and comparing it with the barring rate. If the generated value is not larger than the barring rate, the GF device is activated to start its transmission by randomly selecting one TPL and one RB in the power map. Otherwise, the device access is barred and it has to wait for a fixed barring period for the next attempt.
\end{itemize}
The computation complexity of the user barring algorithm is $\mathcal{O}(\Lambda M)$, where $\Lambda$ is the fixed barring period~\cite{WYu2020}. Apparently, this complexity does not rely on the total number of devices in the network or the actual traffic load at present. Therefore, it is a suitable approach to mitigate the bursty arrivals in MTNOMA-mMTC networks when a massive number of devices get activated in an instant.

After applying user barring techniques, the connectivity of GF devices can be stabilized. Let us consider an extreme scenario where the intra-NOMA interference caused by the access of all GF devices to one RB is smaller than the minimum tolerable interference of all GB devices. This extreme scenario guarantees the QoS of all GB devices, but requires strict constraints on the GF devices' transmit powers. Since the GB connection is ensured under this case, we shall only focus on the AAR of GF devices. Fig. \ref{Fig6a} illustrates the connectivity comparison between a MTNOMA-mMTC network with user barring and that without user barring. One can notice that with user baring techniques, the AAR always remains close to the maximum average throughput (the dash horizontal line). The fluctuations of the AAR curve is due to the randomness of devices' active states and resource selections in each time slot. The gap between the AAR curve with user barring and the theoretical maximum AAR is due to the load estimate errors. If the load estimate is absolutely accurate (which is impossible in practice if there is no extra information exchange between the BS and all devices), the achieved AAR can converge to the theoretical maximum. Then, it means that in the long run, the MTNOMA-mMTC network always operate at the optimal performance.  Fig. \ref{Fig6a} shows that without user baring techniques, the AAR first increases in the light load region and then dramatically decreases when the network load becomes heavy. Fig. \ref{Fig6b} shows that with user barring employed, the average system load in a MTNOMA-mMTC network gradually increases and remains close to optimal load when the number of IoT devices keeps increasing.

\section{Conclusions and Future Challenges}
In this work, we have analyzed the basics of MTNOMA-mMTC networks, which has demonstrated to outperform the traditional OMA-mMTC networks in terms of connectivity and spectral efficiency. A novel power map has been created to generate a practical framework for study the proposed networks. With the aid of semi-GF transmissions, the connectivity has around $40 \%$ max gain than the GB transmissions and $360 \%$ max gain than the GF transmissions under the case with $10$ RBs and $2$ RPLs. For stabilizing the connectivity gain under the scenario with a large number of GF devices, user barring techniques have been applied, which achieves $32$ times higher AAR when there are $200$ devices in the network, given $10$ RBs and $4$ RPLs. In addition, there are still several promising research directions of MTNOMA-mMTC networks, which are listed here:

\begin{itemize}
  \item \textbf{A New Information-Theoretic Framework:} The current information-theoretic framework for mMTC with finite blocklength is based on Gaussian channels. Since PD-NOMA is sensitive to distance-dependent path loss, non-Gaussian channels should be considered. To this end, a tractable expression for non-Gaussian channel capacity with short-packet communications should be acquired via channel coding theory and spatial-domain information theory, which changes the traditional evaluation metric with the unit bit/s to packet/s. Based on this novel expression, a new information-theoretic framework including the active state and available TPLs needs to be proposed.
  \item \textbf{A Unified Spatial Model and ML Structure for Power Map Design:} Note that the efficiency of power maps is mainly dependent on the accuracy of the used mathematical framework. A unified spatial model based on stochastic geometry is necessary to characterize either dense or sparse network environments. After that, the region division can be realized in a heterogeneous scenario. Moreover, the current ML structure for the PP design is based on DRL, which is a model-free ML approach. Future research direction needs to focus on the specific modification of the ML structure to decouple not closely correlated parameters and reduce the computation complexity.
  \item \textbf{QoS-based Semi-GF Transmission Design:} The conventional SIC order for semi-GF transmission is based on the strength order of channel gains. However, in mMTC, the QoS of different IoT devices are various. Some of them need a small data rate but low latency. If these devices have a large channel gain, their information will be decoded later than desire. Although the data rate is higher than expected, this transmission is still failed. To solve this problem, the QoS-based SIC is needed for MTNOMA-mMTC with semi-GF transmissions. It is worth noting that if the power of the signal is smaller than that of the interference-plus-noise when utilizing the QoS-based SIC, a repetition code is required to ensure the decoding.
  \item \textbf{Dynamic User Barring Strategy:} The proposed user barring technique optimizes MTNOMA-mMTC networks by considering the perfect SIC and the pre-defined RPLs. In future studies, an enhanced user barring technique can be proposed by considering imperfect SIC and time-varying received powers due to the varying locations and channel conditions of IoT devices. Furthermore, the backoff mechanisms can be integrated into the user barring technique to further mitigate the collisions and efficiently redistribute access attempts of devices through time.
  \item \textbf{MIMO Design for MTNOMA-mMTC Networks:} MIMO is able to provide both orthogonal (e.g., zero-forcing coding) and non-orthogonal (e.g., dirty paper coding) spatial-domain RBs (SD-RBs) for MTNOMA-mMTC networks, which is able to further enhance the spectral efficiency. The orthogonal SD-RBs are capable of offering no less than one NOMA cluster in each PD-RB. Therefore, the user clustering needs to be redesigned according to the antenna gains in these SD-RBs. The non-orthogonal SD-RBs have the same function but they introduce a new inter-beam interference. Due to this new interference, the principles of user clustering and power allocation in single-antenna MTNOMA-mMTC networks need to be changed. Therefore, resource management for MTNOMA-mMTC networks with MIMO systems requires more technical contributions.
\end{itemize}

\bibliographystyle{IEEEtran}

 \end{document}